\documentclass{emulateapj}

\newcommand{\der}{{\rm d}}
\newcommand{\esc}{_{\rm esc}}
\newcommand{\hal}{_{\rm h}}

\newcommand{\coll}{_{\rm coll}}

\newcommand{\p}{_{\rm p}}

\newcommand{\e}{_{\rm e}}

\newcommand{\ra}{{\cal R}_{\rm a}}

\newcommand{\modot}{M$_\odot$\ }

\newcommand{\nbody}{{$N$}-body }

\newcommand{\beq}{\begin{equation}}
\newcommand{\eeq}{\end{equation}}

\newcommand{\rvir}{R}

\shorttitle{Ellipticals as Violently Relaxed Collisionless
Dissipationless Systems} \shortauthors{Salvador-Sol\'e et al.}

\begin{document}


\title{The Density Profile of Local Ellipticals as Violently Relaxed,
  Collisionless, Dissipationless Systems}

\author{Eduard Salvador-Sol\'e$^{1}$\footnote{$^{1}$Departament d'Astronomia i Meteorologia, Institut de 
Ci\`encies del Cosmos (UB-IEEC), associated with the Consejo Superior de Investigaciones Cient\'\i
ficas, Universitat de
Barcelona, Spain; $^{2}$ Departamento de F\'\i sica Te\'orica, C-XI. 
Universidad Aut\'onoma de Madrid, Spain},
Sinue Serra$^{1}$,
Rosa Dom\'\i nguez-Tenreiro$^{2}$,
and Alberto Manrique$^{1}$
}
\noaffiliation


\begin{abstract}
In a series of recent papers, a new formalism has been developed that
explains the inner structure of dark matter halos as collisionless,
dissipationless systems assembled through mergers and accretion at the
typical cosmological rate. Nearby ellipticals are also collisionless,
dissipationless systems assembling their mass through mergers, but
contrarily to the former structures they do not continuously accrete
external matter because they are shielded by their host halos. Here
we explore the idea that the infall of their own matter ejected within
the halo on the occasion of a violent merger can play a role similar
to external accretion in halos. The predicted stellar mass density
profile fits the observed one, and the empirical total mass density
profile is also recovered.
\end{abstract}



\section{INTRODUCTION}\label{intro}

Nowadays there is increasing evidence about the important role of major,
non-dissipative mergers (Toomre \& Toomre 1972) in the mass
assembly of elliptical galaxies (Conselice 2003; Bell et al.~2005) at
intermediate or low-redshifts, the formation of their stellar
populations and the dissipative processes previous to it having
occured mainly at high-$z$ (Faber et al. 2005; de Lucia et al. 2005;
Dom\'\i nguez-Tenreiro et al.~2006). The problem with dissipationless
major mergers is that there is no known analytical treatment for the
violent relaxation \citep{LyB67} they lead to.

Elliptical galaxies are similar to cold dark matter (CDM) halos in
many respects. Apart from the similar role played by major,
non-dissipative mergers in their mass assembly, their mass, velocity
dispersion, and length scales show systematic regularities and
correlations that can be related with each other (Bernardi et
al.~2003; Graham et al.~2006; O\~norbe et al.~2007).  Likewise, their
3D density profiles are all well fit (Navarro et al.~2004; Merritt et
al.~2005; Merritt et al.~2006; O\~norbe et al.~2007) by the Einasto
(Einasto \& Haud 1969) law or the Prugniel-Simien (1997) approximate
analytical inversion of the projected S\'ersic (1968) law.  These
coincidences suggests that these profiles could have been shaped by
similar physical processes.

In a series of recent papers, a new formalism has been developed
(Salvador-Sol\'e et al.~2007, hereafter SMGH and references therein)
that explains the inner structure of halos from the very
collisionless, dissipationless nature of CDM, and which predicts halo
structural and kinematic properties that are in good agreement with
the results of \nbody simulations (Gonz\'alez-Casado et
al.~2007). Thus, it is natural to consider whether this formalism can
also explain the inner structure of ellipticals.

But things are not that simple. Halos are permanently accreting
external matter, while ellipticals are not. As numerical simulations
show (e.g., Dom\'inguez-Tenreiro et al. in preparation), matter
falling into the halo is too energetic to stick to the center, and it
is deposited at the edge of the system. Only clumps massive enough are
braked by dynamical friction and spiral down until merging with the
central galaxy. Therefore, apart from such discrete (minor and
major) mergers there is no continuous accretion into central
ellipticals.

This is a crucial difference in the context of the SMGH model because
in that model the structure of halos is determined precisely by their
continuous accretion. Indeed, the profile at the edge of a halo of any
extensive property is set by the rates of infalling and rebounding
matter, which do not depend on the particular inner mass distribution
but only on the current accretion rate. As CDM is collisionless, the
spatial distribution of any property is necessarily at all derivative
orders, implying that the whole respective inner profile adapts to the
external one defined by accretion. As CDM is in addition
dissipationless, the steady inner region remains unaltered during
accretion, causing the profiles to grow from the inside out. This
specific growth allows us to infer the profile of any extensive
property from its typical rate of increase by accretion (see below for
the case of the density profile).

Although local ellipticals do not accrete external matter, for a short
interval after a merger they do collect the (bound) matter ejected in
that violent event. The aim of the present Letter is to explore the
possibility that such an infall of ejecta can play a similar role in
ellipticals as that plaied by cosmological accretion in dark halos.

\section{DYNAMICS OF EJECTA}\label{acc}

When two galaxies pass each other close enough to be tidally
disrupted their content is ejected in all directions. As the
disruption takes place when the merging galaxies are orbiting around
the center of mass, the typical radius $R\p$ of the ejection region is
larger than the half-mass radii $r\e$ of the progenitors.  The
particle velocities in the disrupted system should be approximately
normally distributed,
\beq
f({\bf v})\propto \;\exp\left[-\frac{3v^2}{2\sigma\p^2}\right]\,,
\label{veldis}
\eeq
with both the proportionality factor and the characteristic 3D
velocity dispersion $\sigma\p$ independent of the particle mass (the
acceleration undergone by particles does not depend on it).

Particles ejected at larger radial velocities expand more rapidly
than those ejected at smaller ones. Furthermore, for a given
velocity, particles ejected from inner regions experience a smaller
gravitational pull and expand more rapidly than those ejected from
outer regions. Therefore, ejecta tend to segregate into concentric shells
with outwards increasing radial velocities. Most of these shells
follow bound orbits. They reach the apocenter and fall back, cross
the central region and rebound to a somewhat smaller turn-around
radius owing to the crossing with shells falling in for the first
time, and so forth. The result of this chaotic motion is a central
relaxed object into which bound ejecta continue to fall and rebound
for some time.

These dynamics greatly resemble those of the cosmological evolution of
density perturbations, with the difference that in the present case,
apart from the shell-crossing between infalling and rebounding layers,
there is some additional shell-crossing owing to the finite size of
the ejection region as there tends to be segregation of particles
according to radial velocities. In the present Letter, we do not
consider these finite size effects. We focus on the density profile of
the steady object that would emerge were all particles ejected from
the same typical radius $R\p$, and the above mentioned segregation
would be instantaneously and perfectly achieved. In these
circumstances, some known results for the collapse of spherical
density perturbations hold.

If rebounding shells were suppressed, there would be no shell-crossing
(see below) and the stellar mass $M(v)$ inside the shell with positive
initial radial velocity $v$ would be constant and equal to the total
mass of shells with smaller velocities. Taking into account equation
(\ref{veldis}), after integrating over the tangential velocity
components, we obtain for a radial velocity $v$
\beq M(v)=A\!\int_0^{v}\! \der \tilde v\,\exp\!\left(-\frac{3\tilde
v^2}{2\sigma\p^2}\right) =\sqrt{\frac{\pi}{6}}A\sigma\p\,{\rm erf}
\!\left(\frac{\sqrt{3}v}{\sqrt{2}\sigma\p}\right)\!.
\label{Mvi}
\eeq
Similarly, the constant energy of stars inside that shell (for the
system truncated at its radius and the potential origin at infinity)
is
\begin{eqnarray}
E(v)=\!A\!\int_0^{v}\! \der \tilde v\!
\left[\frac{\tilde v^2}{2}+\frac{\sigma\p^2}{3}-\frac{GM(\tilde v)}
{\eta R\p}+\Phi\hal(R\p)\right]\nonumber\\
\times\exp\!\left(\!-\frac{3\tilde v^2}{2\sigma\p^2}\!\right).
~~~~~~~~~~~~~~~~~~~~~~~~~~~~~~~~~~~
\label{Evi}
\end{eqnarray}
The integrals in equations (\ref{Mvi}) and (\ref{Evi}) extend over
\emph{positive} radial velocities only because particles with negative
initial radial velocity, $-v$, cross the system, and join the shell
with $v$, the time required to do so being one of the finite size
effects we ignore. By taking $v$ equal to infinity in these equations,
we arrive at
\beq 
\sigma\p^2=\frac{2E\p}{M\p}+\frac{GM\p}{\eta R\p}-2\Phi\hal(R\p)\,
\quad{\rm and} \quad
A=\sqrt{\frac{6}{\pi}}\frac{M\p}{\sigma\p}\,,
\label{constants}
\eeq
relating the velocity dispersion $\sigma\p$ and the normalization
constant $A$, which is equal to the proportionality factor in equation
(\ref{veldis}) times the average particle mass\footnote{The particle
mass factorizes because the velocity distribution function
(\ref{veldis}) does not depend on it.}, to the typical ejection radius
$R\p$ and the total stellar mass $M\p$, energy $E\p$, and baryon mass
fraction $\eta$ of ejecta. In equations (\ref{Mvi}) and (\ref{Evi}),
$G$ is the gravitational constant and
\begin{eqnarray} 
\Phi\hal(r)=-4\pi G\rho_0\,r^2\hal\,n\hal
P\left[3n\hal,(r/r\hal)^{1/n\hal}\right]~~~~~~~~~~~\nonumber\\
\times\left(\frac{r\hal}{r}\Gamma(3n\hal)+\Gamma(2n\hal)\left\{1-P\left[2n\hal,(r/r\hal)^{1/n\hal}\right]\right\}\right)
\label{Phihal}
\end{eqnarray}
is the potential of the steady dark halo, endowed with an Einasto
density profile (SMGH),
\beq 
\rho(r)=\rho_0\,\exp\left[-\left(\frac{r}{r\hal}\right)^{1/n\hal}\right]\,,
\label{rhohal}
\eeq
with two independent shape parameters, say, $n\hal$ and $r\hal$ for
one fixed total mass $M\hal$. In equation (\ref{Phihal}), $\Gamma(x)$
is the gamma function and $P(a,x)$ is defined as
\beq 
P(a,x)=\frac{1}{\Gamma(a)}\int_0^x \der \xi \exp(-\xi)\, \xi^{a-1}\,.
\label{P}
\eeq
In deriving equation (\ref{Evi}), we have taken into account that, as
the CDM distribution in the steady halo adapts to external accretion,
the void left by the ejected dark matter will be rapidly filled again,
causing the net amount of ejected CDM to decrease. For this reason,
$\eta$ may be substantially larger than the original baryon mass
fraction $\eta\p$ within $R\p$, the relation between the values being
\beq 
\eta=\frac{M\p}{\eta\p^{-1}M\p-M\hal(R\p)}\,,
\label{etap}
\eeq
where $M\p$ is the total ejected stellar mass and 
\beq
M\hal(r)= 4\pi\rho_0 r\hal^3 n\hal\, \Gamma(3n\hal)P\!\left[3n\hal,(r/r\hal)^{1/n\hal}\right]
\label{tcollexp}
\eeq
is the steady halo mass inside $r$ for the Einasto density profile
(eq.~[\ref{rhohal}]).

As mentioned above, rebounding and shell-crossing lead to a steady
state object, which extends, at any given time $t$, out to the
limiting radius $R(t)$. This radius can be estimated from the virial
relation $W(t)=2E(t)$, where $W(t)$ is the potential energy of the relaxed
stellar system, truncated at $R(t)$ and assumed to have a uniform mass
distribution, and $E(t)$ the corresponding conserved (in the previous
model with no shell-crossing) total energy. Estimating $E(t)$ is at
turn-around and assuming a uniform mass distribution, in the
absence of any external gravitational potential, leads to $R(t)$ equal to
half the turn-around radius \citep{GG72}. In the present case, $E(t)$
is given directly by equation (\ref{Evi}) where $v$ is equal to the
initial radial velocity of the shell collapsing at $t$. Dividing that
virial relation by the corresponding inner stellar mass
(eq.~[\ref{Mvi}]), we are led to the following implicit equation for
$R(v)$
\beq 
-\frac{GM(v)}{\eta R(v)}+\Phi\hal[R(v)]=v^2\!+\!\frac{2\sigma\p^2}{3}\!-\!\frac{2GM(v)}{\eta R\p}+
2\Phi\hal(R\p).
\label{rvir}
\eeq 
The initial velocity $v$ of the shell collapsing at $t$ is simply the
inverse of the collapse time, in the model with no shell-crossing, of
the shell that starts with $v$. This collapse time, $t\coll(v)$, can be
obtained by numerical integration of the equation of motion
\beq
\ddot r = -\frac{G\left\{\eta^{-1}M(v)+M\hal[r(t)]\right\}}{r^2}\,,
\label{motion}
\eeq 
then imposing $r(t)$ equal to zero. We have checked that $t\coll(v)$ is an
increasing function of $v$, which proves that there would be no shell-crossing if
the rebounding shells were suppressed.

The density profile of the steady object at $t$ can be obtained in the
way explained in SMGH provided the structure is stable against the
infall of ejecta which warrants its inside-out growth. In other words,
the characteristic time of stellar mass collapse (in the model with no
shell-crossing), inverse of the collapse rate $\ra\equiv \dot M/M$,
must be larger than the crossing time for particles at $R(t)$ for every
collapse time $t$ or, equivalently, for every initial velocity $v$ of
bound shells,
\beq \ra^{-1}(v) >
\left[\frac{\eta \rvir^3(v)}{2GM(v)}\right]^{1/2}\,
\label{taccv}
\eeq
where 
\beq 
\ra(v)=\frac{1}{M(v)}\frac{\der M(v)}{\der
v}\left(\frac{\der t\coll}{\der v}\right)^{-1}\,.
\label{dMt}
\eeq
Condition (\ref{taccv}) is satisfied, so the SMGH model holds.

\begin{figure}
\vspace{-10pt}
\centerline{\includegraphics[width=0.5\textwidth]{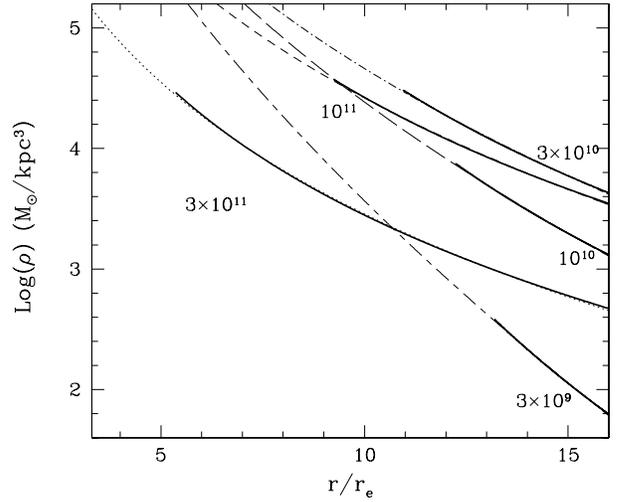}}
\caption{Typical Prugniel-Simien density profiles for ellipticals with
stellar masses equal to $3 \times 10^{9}$ \modot (long-short-dashed
line), $10^{10}$ \modot (dot-dashed line), $3\times 10^{11}$ \modot
(dashed line), and $\times 10^{11}$ \modot (dotted line) according to
Graham et al.~(2006 and references therein) and their best fit by the
theoretical density profile derived here (solid lines). See Table 1
for the values of the respective parameters.}
\vspace{10pt}
\label{1}
\end{figure}
\begin{figure}
\vspace{-10pt}
\centerline{\includegraphics[width=0.5\textwidth]{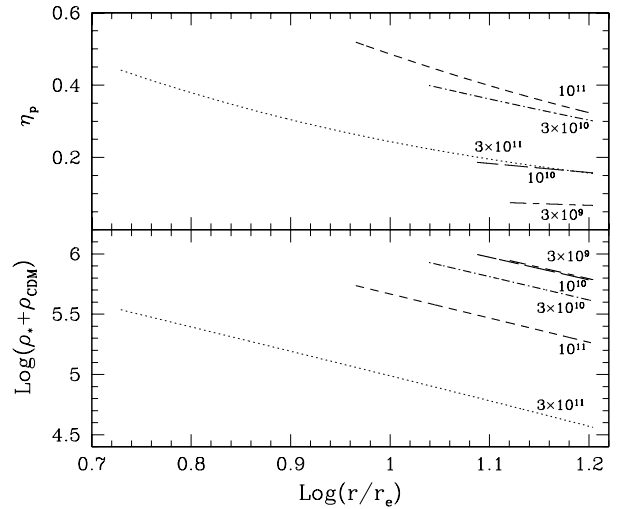}}
\caption{Baryon mass fraction (top panel) and total (baryon or stellar
plus CDM) density profiles (bottom panel) for the theoretical
ellipticals (same symbols) plotted in Figure \ref{1}.}
\vspace{10pt}
\label{2}
\end{figure}

\begin{table*}
\newcommand\T{\rule{0pt}{2.6ex}}
\newcommand\B{\rule[-1.2ex]{0pt}{0pt}}
\begin{tabular}{*{15}c}
\hline
\hline
Galaxy&&\multicolumn{3}{c}{Observed Profile}&&\multicolumn{4}{c}{CDM Halo}&&\multicolumn{3}{c}{Predicted Profile}\T\B\\
\cline{1-1}\cline{3-5}\cline{7-10}\cline{12-14}\B
$M/10^{10}$&&$n$&$r\e$&log $\rho\e$&&$M\hal/10^{12}$&$n\hal$&$r\hal/10^8$&log $\rho_0$&&$R\p$&$\sigma\p$&$\eta\p$\T\\
($M_{\odot}$)&&&(kpc)&$\!\!$$(M_{\odot}$ pc$^{-3}$)&&$\!\!$$(M_{\odot}$)&&(kpc)&$\!\!$$M_{\odot}$ pc$^{-3}$)$\!\!$&&(kpc)&(km s$^{-1}$)&\B\\
\hline
$0.3$&&1.70&0.94&$-0.94$&&$0.20$ (0.05)&6.7 (6.2)&$0.47$ (0.09)&2.4 (2.2)&&24.9 (21.7)&60.0 (36.0)&0.05 (0.16)\T\B\\
$1.0$&&2.34&1.12&$-0.70$&&$0.35$ (0.13)&7.0 (6.6)&$0.25$ (0.71)&2.5 (2.3)&&27.4 (18.6)&92.0 (76.0)&0.14 (0.27)\T\B\\
$3.0$&&3.10&1.57&$-0.72$&&$0.55$ (0.42)&7.2 (7.1)&$0.15$ (0.20)&2.6 (2.5)&&34.3 (34.7)&126 (112)&0.23 (0.28)\T\B\\
$10$&&4.28&3.16&$-1.17$&&$1.40$ (1.40)&7.7 (7.7)&$3.85$ (3.85)&2.9 (2.9)&&58.4 (58.4)&197 (197)&0.29 (0.29)\T\B\\
$30$&&5.72&10.0&$-2.26$&&$4.50$ (4.25)&8.4 (8.4)&$49.6$ (55.4)&3.3 (3.3)&&99.7 (108)&355 (335)&0.27 (0.28)\T\B\\
\hline
\end{tabular}
\caption{}
\label{T1}
\end{table*}

\section{PREDICTED DENSITY PROFILE}\label{dens}

Given the inside-out growth of the steady system, its mass evolves
according to the equation
\beq
M(t)=\int_0^{R(t)} \der r\, 4\pi \rho(r)\, r^2\,, 
\label{mass}
\eeq
where the density $\rho$ is independent of time. Differentiating
equation (\ref{mass}) leads to the density profile $\rho[R(t)]$
in terms of the time derivatives of both the collapsing mass $M(t)$
and the virial radius $R(t)$. More simply, by expressing $M(t)$ and
$R(t)$ as $M[v(t)]$ and $R[v(t)]$ and differentiating equation
(\ref{mass}) by means of the chain rule, the time derivative of $v$
simplifies out, and, after some algebra, we are led to the following
parametric equations for the density profile we want,
\begin{eqnarray} 
\rho(v)=\frac{M(v)+\eta M\hal[R(v)]}{4\pi R^3(v)}
~~~~~~~~~~~~~~~~~~~~~~~~~~~\nonumber\\
\!\!\!\!\!\!\times\!\left\{1\!-\!\frac{2R(v)}{R\p}
\!\!\left[1\!-\!\sqrt{\frac{\pi}{6}}\frac{\eta R\p\sigma\p
v}{GM\p}\exp\!\left(\frac{3v^2}{2\sigma\p^2}\right)\right]\right\}^{-1}
\label{rhor}
\end{eqnarray}
and $R(v)$ given by equation (\ref{rvir}).

In Figure \ref{1}, we plot the typical 3D density profiles of the
Prugniel-Simien form,
\beq 
\rho(r)=\rho\e \left(\frac{r}{r\e}\right)^{-p}\exp\left\{-d_n
\left[\left(\frac{r}{r\e}\right)^{1/n}-1\right]\right\},
\label{Prugniel}
\eeq
for present day ellipticals of various stellar masses $M$. For
$\rho\e$ equal to the density at the half-mass radius $r\e$, $d_n$ and
$p$ become functions of $n$, approximately given, for $n\ga 0.5$, by
$2n-1/3+0.009876/n$ and $1.0-06097/n+0.05463/n^2$, respectively (see
Merrit et al.~2006). Thus, the profile (\ref{Prugniel}) is fixed by
three independent parameters, reducing to two for one given mass
$M$. The typical values of these parameters have been taken from
Graham et al.~(2006 and references therein); see Table 1.

For each empirical profile, we also plot in Figure \ref{1} its best
fit to the theoretical profile given by equations (\ref{rhor}) and
(\ref{rvir}). As ejecta are rapidly collected by the newborn
elliptical (most bound mass is collected in less than one Gyr), its
typical stellar mass $M$ has been assumed to coincide with the final
asymptotic value,
\begin{eqnarray}
M=M\p-A\int_{v\esc}^{\infty}\der \tilde v
\exp\left(-\frac{3\tilde v^2}{2\sigma\p^2}\right)~~~~~~\nonumber\\
=M\p\,{\rm erf}\left[\frac{\sqrt 3}{\sigma\p}\left(\frac{GM\p}
{\eta R\p}-\Phi\hal(R\p)\right)^{1/2}\right]
\label{MMrel}
\end{eqnarray}
where $v\esc=\sqrt{2GM\p/(\eta R\p)-2\Phi\hal(R\p)}$ is the escape
velocity at $R\p$. Then, the typical total dark halo mass $M\hal$ has
been obtained from $M$ by means of the relation by Shanks et
al.~(2006). According to these authors, the ratio of dark halo to
stellar mass, equal to $\sim 14$ for normal ellipticals (see also
O\~norbe et al.~2007), rapidly increases towards the dwarf mass
end. But the estimate is quite uncertain there. For this reason, we
have also considered the assumption of an ever constant dark halo to
stellar mass ratio equal to 14. Then, the Einasto shape parameters
fixing the CDM halo density profile have been obtained using the SMGH
prescription ignoring any adiabatic contraction. Having fixed all
these values (see Table 1) there are only three free parameters to be
adjusted: $\sigma\p$, $R\p$, and $\eta\p$.

As can be seen from Figure \ref{1}, the theoretical profile yields a
very good fit for very reasonable values of those free parameters (see
Table 1): $\sigma\p$ takes values of the order of (or slightly larger
than) the typical velocity dispersion of the progenitors, $R\p$ takes
values of the order of their limiting radii, and $\eta\p$ of the order
of their total baryon mass fraction, (see Fig.~13 in O\~norbe et
al.~2007).

Figure \ref{2} shows the predicted baryon mass fraction and
total (stellar plus CDM) density profiles, which also agree very well
with empirical data (Koopmans et al.~2006; O\~norbe et
al.~2007). Note, in particular, the isothermal-like behavior of the
total density profiles.

\section{CONCLUSIONS}

These results seem to confirm that the structure of nearby ellipticals
(and of simulated non-accreting CDM halos; see Hansen et al.~2005) is
set in major mergers (minor mergers let it essentially unaltered), in
the way explained in SMGH through the infall of matter ejected at that
violent event. At this stage, we found no evidence of a different
formation mechanism for dwarf ellipticals compared to normal bright
ones.

In the present Letter, all ejecta were assumed to be thrown from the
same typical radius. This limits severely the minimum radius down to
which the predicted density profile can be calculated to only $\sim
R\p/2$ (exactly the value in the absence of a dark halo; see
eq.~[\ref{rvir}]), corresponding to the virial radius for shells
initially at rest (at turn-around). In a forthcoming paper, we will
apply a more accurate treatment that will allow us to
reach the galaxy center. In that paper, we will aso deal with the
kinematics of ellipticals.

\acknowledgments 

This work was supported by the Spanish DGES grant AYA2006-15492 and by
the regional government of Madrid through the ASTROCAM Astrophysics
network (S-0505/ESP-0237).  SS benefited from a grant from
the Institut d'Estudis Espacials de Catalunya.



\end{document}